\documentclass[%
reprint,
superscriptaddress,
showpacs,
showkeys,
amsmath,amssymb,
aps,
prl,
]{revtex4-1}

\usepackage{graphicx}
\usepackage{dcolumn}
\usepackage{bm}
\usepackage{upgreek}
\usepackage{gensymb}
\usepackage{nicefrac}%
\usepackage[dvipsnames]{xcolor}
\usepackage{ulem}

\usepackage[pdftex,
            pdfusetitle,
            pdfauthor={Axel Huebl, Martin Rehwald, Lieselotte Obst-Huebl, Tim Ziegler, Marco Garten, Rene Widera, Karl Zeil, Thomas E. Cowan, Michael Bussmann, Ulrich Schramm, Thomas Kluge},
            pdfsubject={scientific paper on theory of laser-ion acceleration},
            pdfkeywords={LPA, laser-ion acceleration, TNSA, multi-species, cryogenic target, particle-in-cell},
            pdfproducer={LaTeX with hyperref},
            pdfcreator={pdflatex}]{hyperref}

\begin{document}

\preprint{APS/123-QED}

\title{Spectral Control via Multi-Species Effects in PW-Class Laser-Ion Acceleration}%

\author{Axel Huebl}
 \email{a.huebl@hzdr.de}
\affiliation{Helmholtz-Zentrum Dresden - Rossendorf, 01328 Dresden, Germany}%
\affiliation{Technical University Dresden, 01062 Dresden, Germany}%
\author{Martin Rehwald}%
\affiliation{Helmholtz-Zentrum Dresden - Rossendorf, 01328 Dresden, Germany}%
\affiliation{Technical University Dresden, 01062 Dresden, Germany}%
\author{Lieselotte Obst-Huebl}%
\affiliation{Helmholtz-Zentrum Dresden - Rossendorf, 01328 Dresden, Germany}%
\affiliation{Technical University Dresden, 01062 Dresden, Germany}%
\author{Tim Ziegler}%
\affiliation{Helmholtz-Zentrum Dresden - Rossendorf, 01328 Dresden, Germany}%
\affiliation{Technical University Dresden, 01062 Dresden, Germany}%
\author{Marco Garten}%
\affiliation{Helmholtz-Zentrum Dresden - Rossendorf, 01328 Dresden, Germany}%
\affiliation{Technical University Dresden, 01062 Dresden, Germany}%
 \author{Ren\'e Widera}%
\affiliation{Helmholtz-Zentrum Dresden - Rossendorf, 01328 Dresden, Germany}%
\author{Karl Zeil}%
\affiliation{Helmholtz-Zentrum Dresden - Rossendorf, 01328 Dresden, Germany}%
\author{Thomas E. Cowan}%
\affiliation{Helmholtz-Zentrum Dresden - Rossendorf, 01328 Dresden, Germany}%
\affiliation{Technical University Dresden, 01062 Dresden, Germany}%
\author{Michael Bussmann}%
\affiliation{Helmholtz-Zentrum Dresden - Rossendorf, 01328 Dresden, Germany}%
\author{Ulrich Schramm}%
\affiliation{Helmholtz-Zentrum Dresden - Rossendorf, 01328 Dresden, Germany}%
\affiliation{Technical University Dresden, 01062 Dresden, Germany}%
\author{Thomas Kluge}%
\affiliation{Helmholtz-Zentrum Dresden - Rossendorf, 01328 Dresden, Germany}%
\date{\today}

\begin{abstract}
Laser-ion acceleration with ultra-short pulse, PW-class lasers is dominated by non-thermal, intra-pulse plasma dynamics.
The presence of multiple ion species or multiple charge states in targets leads to characteristic modulations and even mono-energetic features, depending on the choice of target material.
As spectral signatures of generated ion beams are frequently used to characterize underlying acceleration mechanisms, thermal, multi-fluid descriptions require a revision for predictive capabilities and control in next-generation particle beam sources.
We present an analytical model with explicit inter-species interactions, supported by extensive ab initio simulations.
This enables us to derive important ensemble properties from the spectral distribution resulting from those multi-species effects for arbitrary mixtures.
We further propose a potential experimental implementation with a novel cryogenic target, delivering jets with variable mixtures of hydrogen and deuterium.
Free from contaminants and without strong influence of hardly controllable processes such as ionization dynamics, this would allow a systematic realization of our predictions for the multi-species effect. 
\end{abstract}

\maketitle

High-repetition-rate ultra-short pulse laser driven ion sources receive increasing attention due to their potential as compact particle accelerators\cite{Fuchs2006Laser-drivenIncrease,Wilks2001EnergeticInteractions,Zeil2010TheAcceleration}.
With the ascent of ultra-short pulse (\textless\,$100$\,fs), ultra-high intensity lasers, traditional assumptions for ambipolar ion acceleration in thermal, fluid-like conditions need revision, due to the increasing influence of intra-pulse, pre-thermal dynamics\cite{Gurevich1973IonPlasma,Denavit1979CollisionlessVacuum,Mora2003PlasmaVacuum,Zeil2012DirectAcceleration,Kluge2011ElectronSolids}.
Ion energy spectra are all the more a central observable for experiments and modeling efforts alike, as their shape and maximum energy are characteristic for acceleration mechanisms at given laser and target conditions\cite{Ditmire1997High-energyClusters,Hegelich2002MeVFoils,Robinson2006EffectInteractions,Steinke2013StableRegime,Kar2012IonPressure}.
Also, energy spectra are well accessible experimentally and controlling the spectral shape is important for applications\cite{Sokollik2008TransientDeflectometry,Quinn2009Laser-DrivenSurfaces,Masood2014AProtons}.

Assuming a single ion charge-state in a target, typical target normal sheath acceleration (TNSA) ion spectra follow an exponential distribution\cite{Wilks2001EnergeticInteractions,Mora2003PlasmaVacuum} while light-sail radiation pressure acceleration (RPA) signatures are predicted with a clear spectral gap, and a single quasi mono-energetic ion bunch is produced from thin enough targets\cite{Macchi2009LightReexamined}.
Ion acceleration from thicker targets may have contributions from hole boring RPA at the target front, collisionless shock acceleration, and subsequent rear-side TNSA\cite{Macchi2013IonInteraction}.
The resulting energy spectra feature superelevated dips or even separated quasi mono-energetic proton features\cite{Qiao2012DominanceA2}.
For thin target (surface) layers or transversely small targets, mass-limitation can equally cause quasi-monoenergetic ion spectra independent of the acceleration mechanism\cite{BulanovSV2002GenerationRadiation,Hegelich2002MeVFoils,Esirkepov2002ProposedBeams,Robinson2006EffectInteractions,Pfotenhauer2008SpectralBeams,Zeil2014RobustTargets}, by spatial confinement of the proton source volume to the region responsible for the high-energy tail.

So-called multi-species effects are the result of the ion-ion interaction of varying charge-to-mass constituents during the acceleration phase\cite{Joshi1979QuantitativePlasmas,Begay1982AccelerationTheory,Allen2003ProtonSimulation,Kemp2005MultispeciesDroplets,Psikal2008IonTargets}.
A momentum exchange arises when two such ion species co-propagate during (or after) acceleration.
For the same accelerating fields, "lighter" ($q/m$) species gain higher velocity, leading to charge separation between both expansion fronts.
Additionally, electro-static repulsion between both species establishes shielding of the heavier species' ion front inside the rear electron sheath, transferring momentum to mid-energy light ions of the same phase-space region.
Those light ions are promoted to higher energies, usually without changing their maximum energy at spectral cutoff. \\
In a typical proton acceleration experiment, multiple ion species are present in a thin hydro-carbon contamination layer that inherently covers the target surface.
In this case, or when the foil itself is thin enough, light ions quickly outrun the heavier species' expansion front without significant momentum transfer.
This changes with uniformly composed targets.
Resulting ion energy spectra exhibit a strong dip of light ions at the cutoff energy of the heavier species (per nucleon) with light ions being accumulated at higher energies\cite{Tikhonchuk2005IonFoils,Robinson2006EffectInteractions,Psikal2008IonTargets}.
This effect can in turn be combined with a transverse target mass-limitation, enhancing the energies of a significant fraction of target protons\cite{Kluge2010EnhancedFoils,Hilz2018IsolatedPulse}.
Those multi-species effects were quantified using a two-fluid description, e.g. in \cite{Gurevich1973IonPlasma,Tikhonchuk2005IonFoils,Brantov2006Quasi-mono-energeticPulse}, assuming large ratios between charge-to-mass and density of the two ion species.

In this paper, we systematically characterize the multi-species effect for ultra-short pulse laser interaction with intensities up to the PW-regime.
An analytical model predicting the position of spectral modulation and momentum transfer is presented.
Our model is derived for arbitrary density and charge compositions, extending previous models based on rarefaction wave solutions \cite{Tikhonchuk2005IonFoils,Brantov2006Quasi-mono-energeticPulse} with explicit, scalable inter-species interactions.
For a potential experimental realization, we perform a numerical study in foil-like geometry with density and geometric parameters applicable to a novel target system operated with cryogenic gases\cite{Garcia2014ContinuousHydrogen,Margarone2016ProtonRibbon,Gode2017RelativisticInteractions,Kraft2018FirstTarget,Obst2017EfficientJets,Obst-Huebl2018All-opticalProfiles,Scott2018DualTargets}.
Contrary to composite targets such as plastics, a homogeneously mixed cryogenic target allows on-demand tunability of multiple ion species and ratios at solid density\cite{Kim2018DevelopmentExperiments}. 

In particular, hydrogen and hydrogen-deuterium (H-D) targets provide the following near-ideal starting conditions:
First of all, only two possible charge states $f_i(0, +1)$ exist in H-D plasmas, yet two ion species with differing charge-to-mass ratio are present for multi-species effect studies.
Potential molecular residues can be omitted at given laser intensities.
The operation principle of cryogenic jet targets inhibits the growth of surface contamination layers.
Second, with well-known atomic physics of the system and a density of 30 critical densities, a significant portion of the target can be assumed as pre-ionized under realistic laser temporal contrast.
Cryogenic H-D jets can be produced with fine-grained control of the mixing ratio while keeping the accumulated ion density constant.

In the classical TNSA picture, the front of an ion species expands with $v\sim Z/M$~\cite{Mora2003PlasmaVacuum}.
Under pre-thermal expansion, the resulting distance between two species' ion fronts is small compared to the screening scale length at these fronts%
\footnote{The screening length corresponds to the classical Debye length in a thermalized plasma.
We checked the validity of our assumption in our PIC simulations.
E.g., for $a_0=16$ and medium deuterium concentration of $d=0.5$, the screening length at the de-mixed ion front was measured to be increasingly larger than $0.5\,\upmu$m while the distance between the ion fronts was below $1.0\,\upmu$m up to $t^\mathrm{peak}\approx 1.7\,\tau_\mathrm{laser}$.}.
Exemplified for hydrogen as light species and heavier deuterium, one can solve Gauss's law $ \partial E / \partial y = \rho / \epsilon_0$ for hydrogen ions in front of the spatial deuterium cutoff (ctf), shielding part of the accelerating field.
The rear-side evolution of hydrogen density $n_\mathrm{H}(y,t) = n_{\mathrm{0,H}} \exp{\left(-\nicefrac{y}{c_{\mathrm{s,H}} t} - 1\right)}$ and its electro-static influence upon deuterium ions at cutoff are given as%
\begin{align}
    \Delta E^\mathrm{ctf}_\mathrm{D}(t) &= \frac{Z_\mathrm{H} q_\mathrm{e}}{\epsilon_0} \int^{y^\mathrm{ctf}_\mathrm{H}}_{y^\mathrm{ctf}_\mathrm{D}} n_\mathrm{H} \cdot \mathrm{d}y.\label{eq:dEfield}
\end{align}%
Relevant are only the protons between deuterium and hydrogen spatial cutoff at $y^\mathrm{ctf}_i(t) \cong c_{\mathrm{s},i} t \cdot \left[ 2 \ln{\left(\omega_{\mathrm{p},i} t\right)} + \ln{2} - 3 \right]$ for $\omega_{\mathrm{p},i} t \gg 1$, since charge separation is negligible for small times\cite{Allen1970AWaves,Mora2003PlasmaVacuum}.
$Z_i$ is the ion species' charge, $q_\mathrm{e}$ the (positive) elementary charge, $\epsilon_0$ the dielectric constant, $c_{\mathrm{s},i} \approx \left(Z_i k_\mathrm{B} T_\mathrm{e} m^{-1}_i\right)^{1/2}$ the ion acoustic speed of sound, and $\omega_{\mathrm{p},i}=\left(n_{0,\mathrm{e}} Z^2_i q^2_\mathrm{e} m^{-1}_i \epsilon^{-1}_0\right)^{1/2}$ the prompt ("hot") electron plasma frequency in the rear, scaled for respective ion constants.

We define $d$ as the target deuterium ratio, pure hydrogen targets correspond to $d = 0.0$ and pure deuterium to $d = 1.0$. 
Hence, for the total electron density it follows $n_{0,\mathrm{e}} = Z_\mathrm{D} n_{0,\mathrm{D}} / d = Z_\mathrm{H} n_{0,\mathrm{H}} / (1-d) = Z_\mathrm{H} n_{0,\mathrm{H}} + Z_\mathrm{D} n_{0,\mathrm{D}}$.
Comparing mixed targets to a pure deuterium target, the presence of hydrogen ions leads to a reduced maximum deuterium ion velocity $\Delta v^\mathrm{ctf}_\mathrm{D}$(t) at cutoff, correspondingly shifting the deuterium front $y^\mathrm{ctf}_\mathrm{D}(t)$ by $-\int^t_0 \Delta v^\mathrm{ctf}_\mathrm{D}(t) \mathrm{d}t$.
This shift leads to additional protons overtaking the slowed down deuterium front which in turn changes $v^\mathrm{ctf}_\mathrm{D}(t)$. 
For the sake of simplicity, we neglect this effect and in the following derive an upper estimate of expected deuterium energies.
A straightforward inclusion of this effect would be possible by numerically solving for the resulting integro-differential equation of $\Delta y^\mathrm{ctf}_\mathrm{D}(t)$.

One can integrate the equation of motion $\mathrm{d} p_y(t) / \mathrm{d} t = Z_\mathrm{D} q_\mathrm{e} \cdot \Delta  E^\mathrm{ctf}_\mathrm{D}(t)$, obtaining%
\begin{align}
    \frac{\Delta v^\mathrm{ctf}_\mathrm{D}(t)}{\mathcal{C} \cdot \omega^2_{\mathrm{p,D}}} =& \int^t_{\omega^{-1}_{\mathrm{p,D}}} t \left[\left(\frac{\mathfrak{e}^3}{2\omega^2_{\mathrm{p,D}} t^2}\right)^{\mathfrak{m}_r}-\frac{\mathfrak{e}^3}{2\omega^2_{\mathrm{p,H}} t^2}\right] \mathrm{d}t
\end{align}%
with $\mathcal{C} = (1-d) \cdot Z^{-1}_\mathrm{D} c_{\mathrm{s,H}} / \mathfrak{e}$
($\mathfrak{e}$ is Euler's number), $\mathfrak{m}^2_r = m_\mathrm{H} m^{-1}_\mathrm{D}$ and $\tau_\mathrm{D}=t \cdot \omega_{\mathrm{p,D}}$.

Finally, the cutoff energy $K^\mathrm{ctf}_\mathrm{D}$ of the heavier species, which is equal to the spectral position of the modulation in the lighter ion spectrum, can be predicted from a known scaling of the pure $d=1$ case by $K^\mathrm{ctf}_\mathrm{D}(d) = m_\mathrm{D}/2 \left(v^{d=1}_\mathrm{D} - \Delta  v^\mathrm{ctf}_\mathrm{D} \right)^2$ with%
\begin{align}
    \frac{\Delta v^\mathrm{ctf}_\mathrm{D}(\tau_\mathrm{D})}{ \mathcal{C}} =& \frac{\mathfrak{e}^{3\mathfrak{m}_r}}{2^{\mathfrak{m}_r}} \cdot \frac{\tau^{2-2\mathfrak{m}_r}_\mathrm{D}-1}{2-2\mathfrak{m}_r}
    - \frac{Z^2_\mathrm{D}}{Z^2_\mathrm{H}} \frac{\mathfrak{e}^3 \mathfrak{m}^2_r}{2}\cdot \ln{(\tau_\mathrm{D})}. \label{eq:dVD}
\end{align}%
The energy difference is lost to the lighter species, shifting mid-spectrum protons to higher energies than through regular TNSA.
An effective acceleration time should be applied for $\tau_\mathrm{D}$\cite{Fuchs2006Laser-drivenIncrease}.

Following the 1D nature of this model, the spatial density distribution of the hydrogen ions that have already overtaken the deuterium cutoff is not relevant. 
That assumption is valid along the target normal as long as no transverse forces displace a significant amount of charge from the axis.
With increased deuterium content $d$ in the target, the total number of protons drops accordingly.
Deuterium cutoff-energies then shift to higher energies since the shielding of rear fields decreases.
We obtain a description over arbitrary composition ratios $d$ without a divergence of the predicted energies at $d=1$, contrary to
\cite{Tikhonchuk2005IonFoils,Brantov2006Quasi-mono-energeticPulse}.

Our model in eq.~\eqref{eq:dVD} complements theories for cutoff with signatures visible at significantly higher particle flux in the spectrum.
With the position of modulation scaling different with $T_\mathrm{e}$ and $n^\mathrm{rear}_\mathrm{e}$ than the lighter species' cutoff, the average kinetic electron energy in the sheath $T_\mathrm{e}$ and pre-thermal, laser-accelerated electron density $n^\mathrm{rear}_\mathrm{e}$ can be determined from the same spectra.

We verify our model with numerical 2D3V PIConGPU simulations\cite{Bussmann2013RadiativeInstability,Huebl2019PIConGPUFixes}.
Operating in a standard regime for many Ti:Sa ultra-high intensity short pulse laser systems, the explored parameters include those available at HZDR's Draco ion-acceleration end-stations for its 150\,TW and 1\,PW stages\cite{Schramm2017FirstDresden}.
The final focusing parabola and spot size on target is assumed
to be $w_{0,I}^{\mathrm{FWHM}} = 3.0 \,\mathrm{\upmu m}$\cite{Obst2017EfficientJets}.
The temporal laser profile is modeled as a Gaussian profile with $\tau_I^{\mathrm{FWHM}} = 30\,\mathrm{fs}$.
For the central wavelength of 800\,nm we normalize to the critical density $n_c = 1.74\cdot 10^{21} \mathrm{\,cm}^{-3}$.

We present numerical studies on a 2\,$\mathrm{\upmu m}$ thick, planar jet, disabling effects of lateral mass-limitation\cite{Obst2017EfficientJets}.
Our modeled initial density profile of $n_{0,\mathrm{e}} = 30\,n_c$ is flat, where a short exponential ramp with $L = 20$\,nm scale-length was added at the front to account for slight hydro-dynamic pre-expansion and increasing numerical robustness.
Peak laser intensities on target are varied in steps of the dimensionless field amplitude $a_0 = 8, 16, 23, 30$ and $42$.
The laser pulse propagates in positive $y$ direction, while the focal position is set to the target center at $4\,\mathrm{\upmu m}$.
For detailed analysis, a radial particle acceptance filter is deployed, similar to realistic experimental pinhole apertures with $\pm 2 \degree$. 
Additionally, particles originating in the first half of the target, $y_0 \le 4 \upmu$m, are tracked as target "front" and all other particles as target "rear" for discussion.
Further numerical parameters of our simulations, open source code, scripts and a sketch of the assumed setup are described in the supplementaries. 

\begin{figure}[ht]
  \centering
  \includegraphics[width=0.48\textwidth]{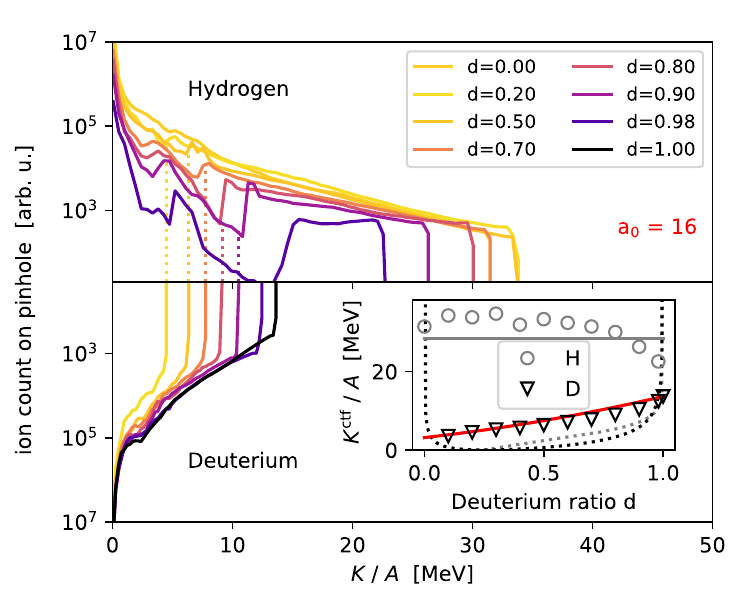}
  \caption{Deuterium \& hydrogen energy spectra for $a_0=16$ for variations in target deuterium ratio $d$.
  The inset displays ion cutoff energies for variations of $d$.
  Our analytical model from eq.~\eqref{eq:dVD} is plotted as red line and a grey line for~\cite{Mora2003PlasmaVacuum} eq.~(22), both with measured average $\langle T_\mathrm{e}\rangle=2.17$\,MeV and $n^\mathrm{rear}_\mathrm{e} = 0.31 \cdot n_\mathrm{0,e}$.
  Dashed lines are previous models for heavy species cutoff (black) and light species peak (grey) in \cite{Tikhonchuk2005IonFoils,Brantov2006Quasi-mono-energeticPulse} eq.~(1).
  }%
  \label{fig:a0_16_energySpectra}
\end{figure}%

\begin{figure}[ht]
  \centering
  \includegraphics[width=0.48\textwidth]{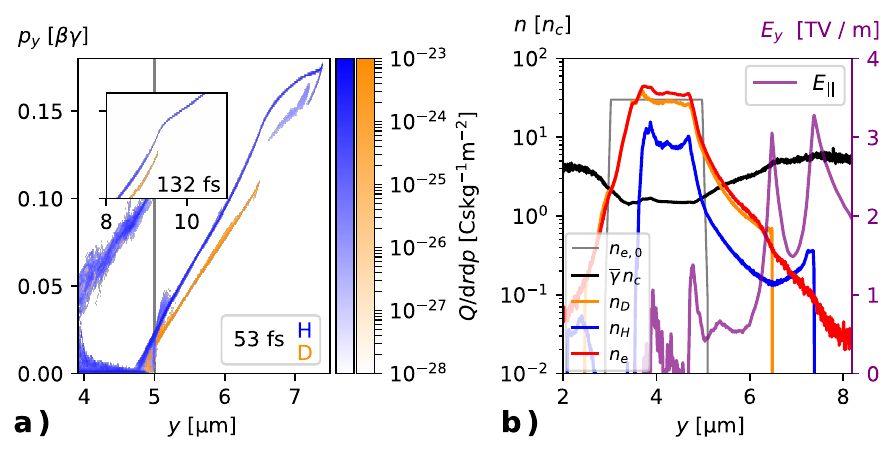}
  \caption{\textbf{a)} Longitudinal phase space for $a_0=16$ for a deuterium ratio $d=80\%$.
  Taken at time $t = 53$\,fs after the peak intensity of the laser pulse reached the target.
  Inset shows the rear ion evolution at $t = 132$\,fs.
  \textbf{b)} Fields shown for same time.
  Clearly visible is the partial shielding of the electron sheath through expanding protons around 7\,$\upmu$m.
  The bump of proton density at its front originates from the initial rear scale length~\cite{Mora2003PlasmaVacuum}.%
  }\label{fig:a0_16_ypy}
\end{figure}

Figure~\ref{fig:a0_16_energySpectra} summarizes the resulting ion spectra for $a_0 = 16$.
An overall exponential shape confirms operation in the TNSA regime\cite{Mora2003PlasmaVacuum}.
The deep modulation of the proton spectra is corresponds to the cutoff energy per nucleon of the heavier deuterium species, the latter we predict with our model from eq.~\eqref{eq:dVD}.
For the same measured average electron kinetic energy $T_e$ and density $n^\mathrm{rear}_\mathrm{e}$ the theoretical predictions for proton cutoff (grey line) and deuterium cutoff (red line) are added.
For $d\xrightarrow{} 1$, quasi-monoenergetic features are formed in the lighter species.

A phase space image in Figure~\ref{fig:a0_16_ypy}a) shows that hydrogen ions expand faster from the target rear side than deuterium, which is consistent with TNSA scalings for the ion front velocity applied in eq.~\eqref{eq:dEfield}.
The deuterium front therefore interacts dominantly with protons in the mid range of the spectrum, which are shifted longitudinally to higher energies due to electro-static repulsion.
A lineout in Figure~\ref{fig:a0_16_ypy}b) shows the resulting modulated proton density and the peaked electric field (purple) at each ion species' front at later time.

Integrating the overall proton signal in "pinhole" acceptance reveals that the on-axis spectrum is indeed solely pushed to higher energies and deuterium-caused "loss" of protons on-axis due to transverse displacement is negligible.
Consistently, the spectral hydrogen accumulation after each dip in Figure~\ref{fig:a0_16_energySpectra} consists of the protons in the dip.
However, by keeping the initial target density constant with mixing ratios $d$, proton counts for mid-spectrum energies do not exceed values of the pure-hydrogen case.

\begin{figure}[ht]
  \centering
  \includegraphics[width=0.48\textwidth]{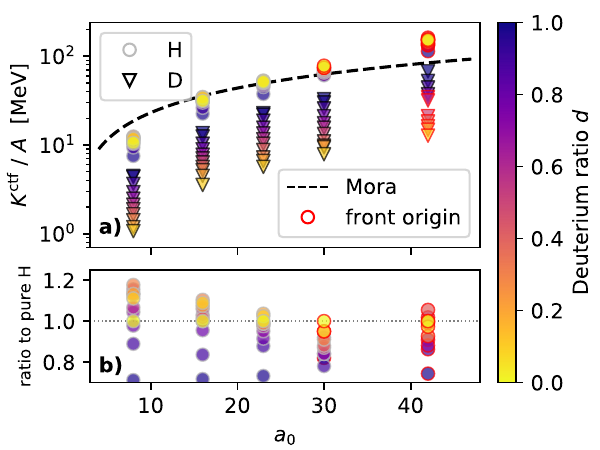}
  \caption{
  \textbf{a)} Deuterium (triangle) and hydrogen (sphere) cutoff energies with varying laser intensity.
  For a given $a_0$, each H-D pair with the same deuterium ratio $d$ (color) corresponds to one simulation.
  Dashed line is a simple estimate for Mora-scaling~\cite{Mora2003PlasmaVacuum} with constant $n^\mathrm{rear}_e=0.35$ and $\langle T_\mathrm{e}\rangle$ as in~\cite{Kluge2011ElectronSolids}.
  \textbf{b)}~Zoomed plot for hydrogen cutoff energies:
  relative energy increase/decrease compared to a pure hydrogen target ($d=0$) under variation of deuterium content $d$.
  The variation of target composition leads to enhancements up to 18\% and reduction of 28\% in hydrogen cutoff energy, respectively.}\label{fig:cutoff_scaling}
\end{figure}

As we vary the laser intensity on target in Figure~\ref{fig:cutoff_scaling}a), the increase of deuterium cutoff energy with deuterium mixing ratio $d$ holds true, even for PW-scale laser intensities.
Front-side ions start to dominate the highest energies for $a_0 \gtrsim 30$, as indicated by red marker edges in figure~\ref{fig:cutoff_scaling}.
As the hole-boring velocity for protons at the target front surface increases with $a_0$~\cite{Robinson2009RelativisticallyPulses}, front surface originating ions are initially faster and hence reach the target rear fields in shorter time.
Rear-side electro-static TNSA fields decay on time scales of the laser pulse length $\tau_\mathrm{laser}$.
Consequently, quick front to rear surface transport, controlled by target thickness, enables hole-boring accelerated protons (kinetic energy $K_\mathrm{HB}=0.26-6.0$\,MeV for the given $a_0$) to undergo further acceleration through rear side TNSA\cite{Qiao2012DominanceA2}.

Calculation of the relativistic skin depth $c / \omega^\mathrm{rear}_{\mathrm{p,e}} = 98$\,nm suggests that a $2\,\upmu$m thick, $30\,n_c$ target remains opaque even for the highest $a_0=42$ (average relativistic gamma factor for electrons is $\gamma^\mathrm{rear}_e = 17.9$).
However, the hole boring front burns through the target before the end of the laser intra-pulse phase for $a_0 \gtrsim 20$, resulting in the observation of transmitted light.
Field lineouts are shown in Fig.~\ref{fig:app_rearFields} in supplementaries.

This study is performed with an overall conserved target free electron density. 
In the TNSA regime with thick targets, deuterium ions only modulate mid-energy protons and the cutoff energy for protons for a constant $a_0$ would be expected to be independent of $d$, hence we plot a constant prediction for Mora-scaling in Figure~\ref{fig:a0_16_energySpectra}.
Intriguingly, our simulations as well as data in previous studies show a monotonically decreasing cutoff energy with proton density\cite{Robinson2006EffectInteractions}.
Furthermore, we observe a slight increase (up to 18\,\%) in proton cutoff energy as compared to pure hydrogen targets, detailed in the linear plot of Figure~\ref{fig:cutoff_scaling}b).
This secondary effect is likely attributed to slightly changed absorption efficiency on the target front, for example due to micro-structuring of the surface, as we measure a change in average electron energy density (refer to supplementary material for details).

In summary, we developed an analytical model for the multi-species effect, predicting the reduced cutoff energy of a heavier ion species and associated position of spectral modulations for arbitrarily mixed, homogeneous composite targets in planar geometry.
We calibrated our model against the cutoff energy of a target consisting purely of the heavier species, which can be either measured, simulated or derived from established theories\cite{Mora2003PlasmaVacuum,Kluge2011ElectronSolids,Schreiber2006AnalyticalPulses}.

Combining our results for spectral modulations with models for proton cutoff energy offers advanced predictive capabilities for properties of the laser-driven electron population.
In particular, average electron kinetic energy and density assumptions can be self-consistently verified against both cutoff and modulation in the same spectra.

We performed 60 systematic ab initio simulations on multi-species effects for a planar, homogeneous cryogenic target, irradiated with PW-scale laser intensities. 
Spectral signatures observed in our simulations are unique to multi-species effects and exhibit striking agreement with analytic predictions.
A deuterium-hydrogen jet target was investigated in view of a methodical experimental realization, as its operation principle provides easy control over the target constituents and inhibits surface contamination, thus limiting the available charge states and simplifying ionization physics.
Thus, this target offers the unique opportunity to study multi-species effects at variable target composition in a well-defined environment.
Our model is however open to in-detail adjustments of individual starting conditions of laser and target parameters in view of many different experimental situations, as well as uncertainties and adjustments of specific numerical models used, e.g. for ionization or collisions.

The presented analytical model in eq.~\eqref{eq:dVD} is not limited to the hydrogen-deuterium jets addressed in this study.
Adaption to other target constituents with different charge-to-mass ratios is straight-forward.
In fact, virtually all ion acceleration schemes at ultra-high laser intensities have relevant contributions from multi-species effects even for pure materials, where different ionization states take the role of the different ion species addressed in this work.

\begin{acknowledgments}
The work has been partially supported by EC Horizon 2020 LASERLAB-EUROPE/LEPP (Contract No. 654148).
This project received funding within the MEPHISTO project (BMBF-F\"orderkennzeichen 01IH16006C).
Computations were performed on the Hemera cluster at Helmholtz-Zentrum Dresden - Rossendorf.
We acknowledge all contributors to the open-source code PIConGPU for enabling our simulations.
We thank all contributors to the SciPy ecosystem for enabling our data analysis.
We thank F. Fiuza, S. G\"ode, Y. Long, and L. Huang for fruitful discussions.
\end{acknowledgments}

%

\onecolumngrid
\newpage
\appendix
\section*{Appendices}
\addcontentsline{toc}{section}{Appendices}
\renewcommand{\thesubsection}{\Alph{subsection}}

\section{\label{app:opendata}Data availability}

Simulation input files, in situ analysis output, scripts and analyzed data that support the figures and other findings of this study are available from:

\begin{center}
    DOI:10.14278/rodare.116 \\
    URL: https://doi.org/10.14278/rodare.116
\end{center}

PIConGPU is open source and all versions, including 0.4.3 used in this paper, are available as download and contributable git repository\cite{Bussmann2013RadiativeInstability,Huebl2019PIConGPUFixes}.

Additional high-resolution, raw HDF5 files using the openPMD standard (DOI:10.5281/zenodo.1167843)
increase simulation output data to 4.7\,TByte and are available from the corresponding author upon reasonable request.

\section{\label{app:simsetup}Simulation Setup Overview}

\begin{figure}[ht]
  \centering
  \includegraphics[width=0.5\textwidth]{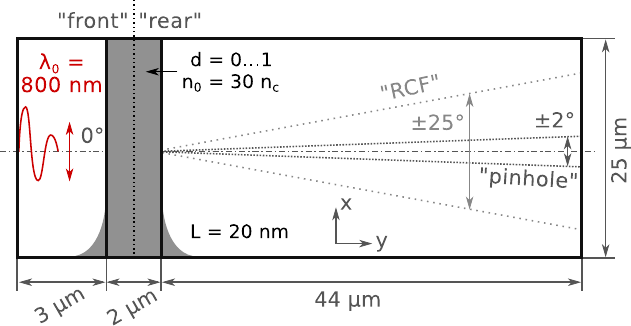}
  \caption{Simulation setup and used naming conventions.
  In situ applied particle filters are modeled according to a synthetic RCF stack or pinhole, e.g. in a Thomson parabola.
  Laser incident angle is zero degrees (from left along $y$) and polarization is parallel to the simulated plane.}\label{fig:setup}
\end{figure}

Figure~\ref{fig:setup} shows the simulation box, with the laser pulse propagating in positive $y$ direction.

The initial density profile is modeled after (all in $\upmu$m):%
\begin{align}
	n_e(y) &= 30\,n_c \cdot 
    \begin{cases}
      \mathrm{e}^{-(y-5)/0.02}, & \text{if}\ y>5\\
      1, & \text{if}\ 3<y<5 \\
      \mathrm{e}^{(y-3)/0.02}, & \text{if}\ y<3
    \end{cases}
              \label{eq:ne}
\end{align}%
Numerical particle-in-cell solvers deployed are Yee Maxwell-solver, optimized Esirkepov current deposition (ZigZag path splitting), randomized in-cell starting positions without temperature, 3rd order (piecewise cubic) particle assignment shape, Boris particle pusher, and weighted trilinear force interpolation.
Collision operators are neglected.

The 2D simulations' spatial resolution is $\Delta x, \Delta y = 3.33\,\mathrm{nm}$ with $\Delta t = 7.85\,\mathrm{as}$ on a grid of 7488x14720 cells.
Equivalently, the central laser wavelength is resolved with 240\,cells and the non-relativistic plasma frequency $\omega_\mathrm{p} \cdot \Delta t = 0.10$.
Per ion species, 20 particles per cell and one electron particle per ion is initialized as pre-ionized macro-particle distribution.
Each simulation computes $0.60\,\mathrm{ps}$ interaction time on 16 GPUs (Nvidia P100) within 1:20-1:45\,hours.

The laser pulse's peak intensity reaches the target after 8190 steps (64.3\,fs).
Energy spectra were compared on final, converged results before particles leave the simulation box, at step 90\,000, 75\,000, 70\,000, 60\,000 or 45\,000, depending on laser intensity.

\section{\label{app:further_spectra}Simulation and Analytical Results for $a_0=8,23,30,42$}

The explicit form of the prediction for the heavier species' cutoff energy in eq.~\eqref{eq:dVD} is:%
\begin{align}
     K^\mathrm{ctf}_\mathrm{D}(d, T_\mathrm{e}) &=
         \frac{m_\mathrm{D}}{2} \cdot \left(
             \sqrt{K^{d=1\mathrm{,ctf}}_\mathrm{D} \cdot \frac{2 A_\mathrm{D}}{m_\mathrm{D}}}
             - \frac{(1 - d)}{\mathfrak{e} \cdot Z_\mathrm{D}}
             \sqrt{\frac{Z_\mathrm{H} k_\mathrm{B} \langle T_\mathrm{e} \rangle}{m_\mathrm{H}}} \cdot
             \left(
                \frac{\mathfrak{e}^{3\mathfrak{m}_r}}{2^{\mathfrak{m}_r}} \cdot
                \frac{\tau^{2-2\mathfrak{m}_r}_\mathrm{D}-1}{2-2\mathfrak{m}_r}
                - \frac{Z^2_\mathrm{D}}{Z^2_\mathrm{H}} \frac{\mathfrak{e}^3 \mathfrak{m}^2_r}{2}\cdot \ln{(\tau_\mathrm{D})}
             \right)
         \right)^2.
\end{align}

All quantities are given in SI,
$\mathfrak{e}$ is Euler's number, $\mathfrak{m}^2_r = m_\mathrm{H} m^{-1}_\mathrm{D}$ and $\tau_\mathrm{D}=t \cdot \omega^{d=1}_\mathrm{p,D} \approx 1.3 \cdot \tau^\mathrm{I,FWHM}_\mathrm{laser} \cdot \omega^{d=1}_\mathrm{p,D}$ are unitless\cite{Fuchs2006Laser-drivenIncrease}.

\begin{figure*}[ht]
  \centering
  \includegraphics[width=\textwidth]{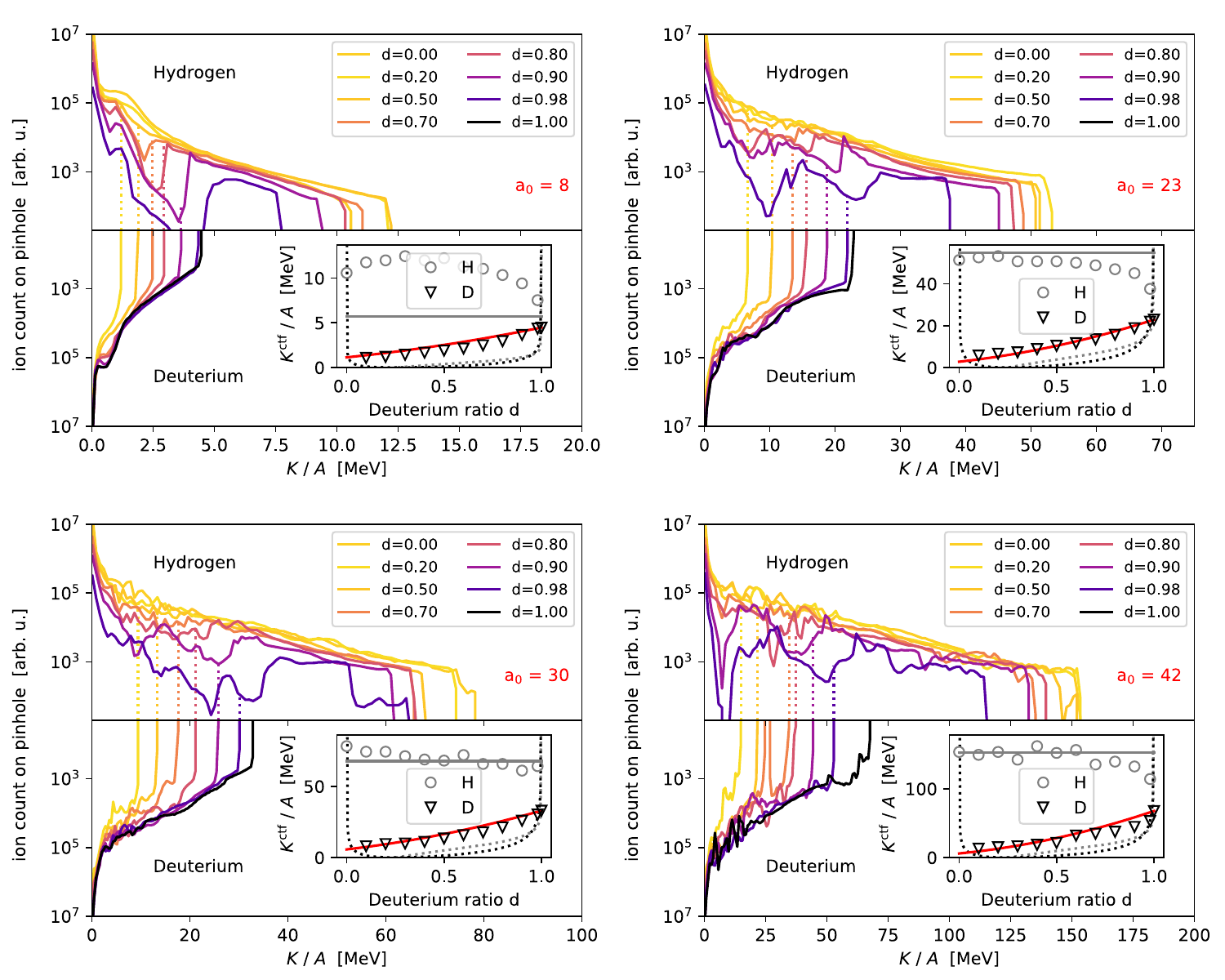}
  \caption{Deuterium and hydrogen energy spectra for laser strengths parameter $a_0=8-42$.
  Individually shown as in Figure~\ref{fig:a0_16_energySpectra} for variations in target Deuterium ratio $d$.
  Ions shown for $\pm$2 degree "pinhole" acceptance.
  The theoretical prediction from eq.~\eqref{eq:dVD} is presented as red solid line and hydrogen cutoff following Mora scaling as grey line~\cite{Mora2003PlasmaVacuum} for $\langle T_\mathrm{e}\rangle=(0.35, 3.52, 4.52, 8.64)$\,MeV and non-perfect absorption into rear electrons $n^\mathrm{rear}_\mathrm{e} = (0.48, 0.44, 0.40, 0.57) \cdot n_\mathrm{0,e}$ for $a_0=(9, 23, 30, 42)$.
  Dashed lines are previous models ($d\approx 1$) for heavy species cutoff (black) and light species peak (grey) from eqs.~\eqref{eq:Dold}, \eqref{eq:Hold}.%
  }%
  \label{fig:a0_8_23_30_40_energySpectra}
\end{figure*}

Replacing $Z_\mathrm{D}, A_\mathrm{D}, m_\mathrm{D}$ for any other heavy ion species or choosing another "light" ion species for the same quantities in $X_\mathrm{H}$ variables should hold true and no assumptions over the abundance or ratio between the two are taken into the derivation.
Note to readers: previous papers \cite{Tikhonchuk2005IonFoils,Brantov2006Quasi-mono-energeticPulse} annotated quantities $X_\mathrm{H}$ for "heavy" and $X_\mathrm{L}$ for "light" ions.
Our choice of $\mathrm{H}$ of the light hydrogen (proton) species is the exact opposite.
Translating \cite{Tikhonchuk2005IonFoils,Brantov2006Quasi-mono-energeticPulse} eq.~(1) in our nomenclature reads%
\begin{align}
    K^\mathrm{ctf}_\mathrm{D} &= Z_\mathrm{D} \langle T_\mathrm{e} \rangle \ln^2\left[4Z_\mathrm{D} d\sqrt{2 A_\mathrm{D} Z_\mathrm{H} / A_\mathrm{H} Z_\mathrm{D}}/(Z_\mathrm{H} (d-1) \mathfrak{e})\right]/2\label{eq:Dold}\\
    K^\mathrm{peak}_\mathrm{H} &= Z_\mathrm{H} \langle T_\mathrm{e} \rangle \ln\left[4Z_\mathrm{D} d\sqrt{2 A_\mathrm{D} Z_\mathrm{H} / A_\mathrm{H} Z_\mathrm{D}}/(Z_\mathrm{H} (d-1) \mathfrak{e})\right].\label{eq:Hold}
\end{align}%
In our model, we derive the cutoff energy $K^\mathrm{ctf}_\mathrm{D}$ of the heavier species.
As the lighter species is shifted to higher energies from this spectral position, a peak is observed at $K^\mathrm{peak}_\mathrm{H}$ for protons.
Previous models described this as in eq.~\eqref{eq:Hold}.
This is the grey dotted line in Figures~\ref{fig:a0_16_energySpectra} and \ref{fig:a0_8_23_30_40_energySpectra}.

Figure~\ref{fig:a0_8_23_30_40_energySpectra} provides additional spectra collection for various simulated laser intensities of the main text.
Averaged kinetic energies $\langle T_\mathrm{e}\rangle$ and $n^\mathrm{rear}_\mathrm{e}$ are a best-effort to measure in situ from simulation data.
Please see assumptions and method in provided scripts.
Alternatively, like previous authors one can also just apply a theoretical scaling for the average kinetic energy, such as \cite{Kluge2011ElectronSolids},
and approximate $n^\mathrm{rear}_\mathrm{e}\approx n_0 \cdot \eta$ via the absorption in Figure~\ref{fig:energyConversion}.

\newpage
\section{\label{app:absorption}Determination of $\langle T_\mathrm{e}\rangle$ and $n^\mathrm{rear}_\mathrm{e}$}

\begin{figure*}[ht]
  \centering
  \includegraphics[width=0.5\textwidth]{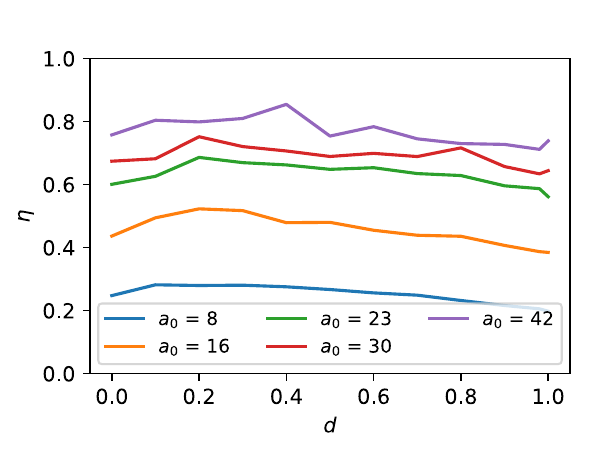}
  \caption{Energy conversion $\eta$ from laser energy to kinetic particle energy over laser strength parameter $a_0$ and relative deuterium content $d$.
  }\label{fig:energyConversion}
\end{figure*}

Figure~\ref{fig:energyConversion} shows the relative ratio of laser energy absorbed by the target.

\begin{figure*}[ht]
  \centering
  \includegraphics[width=0.90\textwidth]{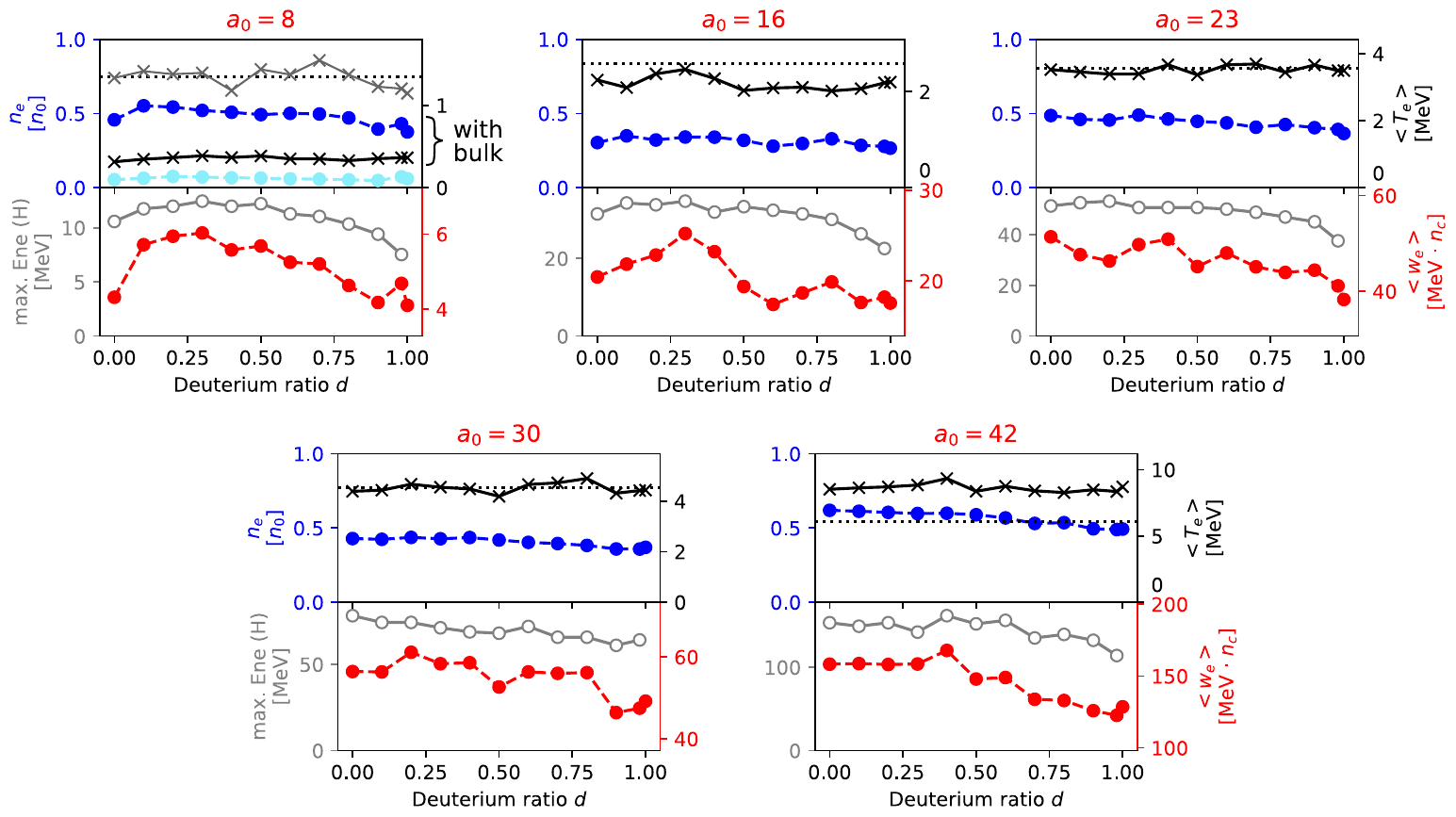}
  \caption{In situ measured parameters for laser accelerated electrons: density ratio to cold target, average kinetic energy, resulting average energy density versus hydrogen cutoff energy.
  The black dashed line is the theoretical scaling from~\cite{Kluge2011ElectronSolids} eq.~(9).
  }\label{fig:Te_ne_measured}
\end{figure*}

In order to estimate the average kinetic energy $\langle T_\mathrm{e}\rangle$ and $n^\mathrm{rear}_\mathrm{e}$ for electrons in simulations, various methods are used by contemporary literature, yet rarely documented.
Either apply one of the theoretical models for $\langle T_\mathrm{e}\rangle$, such as \cite{Kluge2011ElectronSolids}, and assume the all-target absorption $\eta \approx n^\mathrm{rear}_\mathrm{e} / n_{0,\mathrm{e}}$ or try to measure in situ as shown in Figure~\ref{fig:Te_ne_measured}.

In our analysis of electrons in the simulation, we select a spatial region around the laser axis in the second half of the target and measure the prompt, laser-accelerated electrons with relativistic values for $\gamma_\mathrm{e}$.
We measure shortly after the laser peak intensity reaches the target.
The resulting average kinetic energy and density is then corrected for the missing low-energy contribution due to our spectral $\gamma_\mathrm{e}$-cutoff, assuming a Boltzmann distribution, which we verify in electron energy histograms for the assumed "prompt" selection.

We observe derivations from those predictions at $a_0=8$ and $42$.
Comparing theoretical predictions for Hydrogen cutoff energy for $a_0=8$, we realize our selection of prompt electrons underestimates the electron density relevant for acceleration of ions.
We therefore included a fraction of bulk electrons in this specific measurement in order to account for re-heating effects and slow expansion, which is influenced by non-prompt electrons at this low intensity.
Nevertheless, this method of measurement likely underestimates the average kinetic energy of electrons due to crude averaging of both bulk and prompt electron distributions.
Due to burn-through of the target for the $a_0=42$ case, the laser can penetrate the target and electrons can co-propagate with its phase, increasing the average electron energy above the case of a solid, reflecting surface and therefore values exceed the dashed black line, showing Ref.~\cite{Kluge2011ElectronSolids} eq.~(9).

Observing the energy-density of laser accelerated electrons in Figure~\ref{fig:Te_ne_measured} (red data points), a maximum is achieved for a deuterium ratio of about $d=20-50$\,\% for $a_0<20$.
Following \cite{Mora2003PlasmaVacuum}, this translates into a maximum acceleration field and hence maximum proton energies (grey data points).

We therefore conclude that the deuterium ratio in the target modifies the front side absorption of the laser energy into Debye sheath electrons.
Since we do not observe a correlated change in the front side plasma steepness, this change in absorption is likely due to varied micro-structuring of the surface or other 2D effects.

\section{\label{app:rear_fields}Target Rear}

\begin{figure}[ht]
  \centering
  \includegraphics[width=0.80\textwidth]{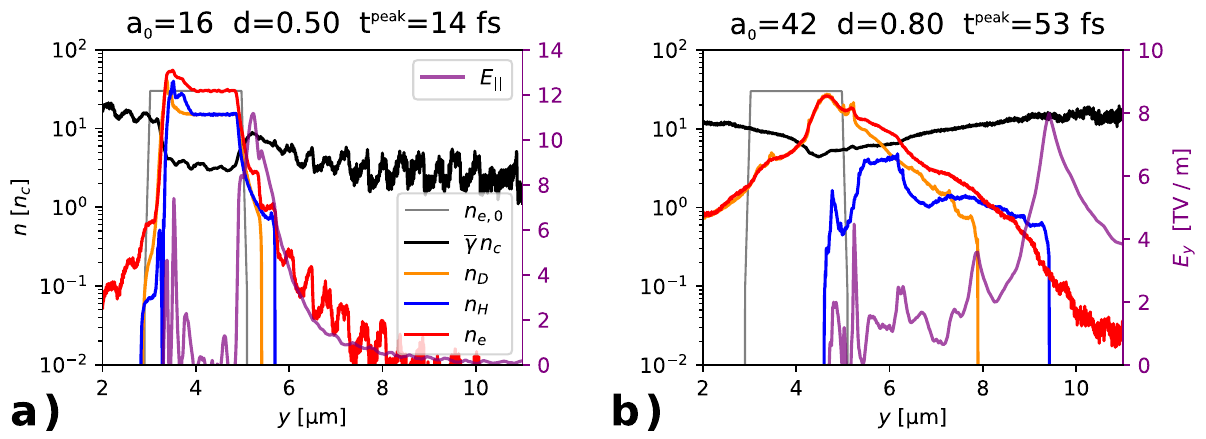}
  \caption{Lineouts of plasma density and longitudinal electric field an early stage of the acceleration.
  \textbf{a)} $a_0=16$ and target deuterium ratio $d = 0.5$.
  \textbf{b)} $a_0=42$ and target deuterium ratio $d = 0.8$.%
  }\label{fig:app_rearFields}
\end{figure}

Figure~\ref{fig:app_rearFields}a) exemplifies the electric field on the target rear.
Clearly visible is the $2\omega_0$ imprinting of the laser frequency onto rear-side, promptly accelerated electrons.
From comparisons between different deuterium ratios $d$, we find that an optimal absorption of laser energy into maximum number of electrons on the target rear ($y>5.0 \upmu$m) and maximum accelerating field can be designed with about 20 - 50\,\% deuterium in the target.
This leads to cutoff energy increase for $a_0 = 8 - 16$ of up to 18\,\% compared to a pure proton target of the same density.

Figure~\ref{fig:app_rearFields}b) demonstrates that the target stays locally opaque even for $a_0 = 42$ during laser interaction: note the section at $y=4-5\,\upmu$m on which the electron density (red) is larger than the relativistic critical density (black).
Nevertheless, plasma filaments are observed breaking up transverse sections of the target enabling burn-though.

\section{\label{app:emission-distribution}Ion Emission Distribution}

\begin{figure*}[ht]
  \centering
  \includegraphics[width=\textwidth]{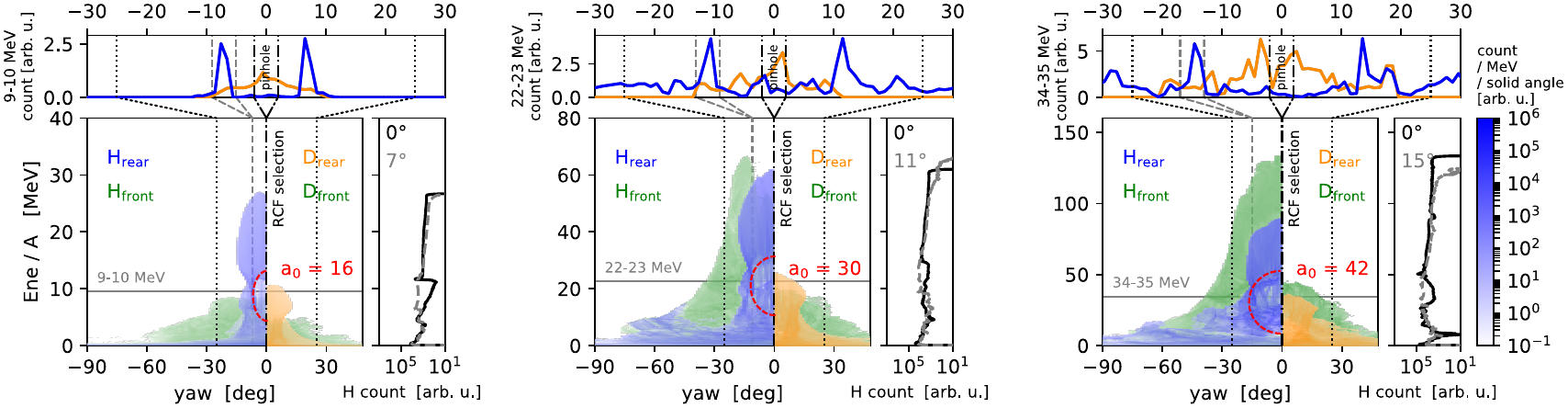}
  \caption{\textbf{Lower plots:} Emission distribution of ions.
  Plotted for laser intensities $a_0=16,30,42$ (left-right) and target deuterium ratio $d=90\%$.
  Left half is hydrogen (blue) and right half deuterium ions (orange) with green denoting front-originating ions in both cases.
  For large enough angle to the laser axis and energy, one can infer that only front-side ions contribute (areas plotted solely in green).
  Rear-side ions are plotted on top of front-side ions.
  \textbf{Upper plots:} A sample lineout, zoomed into a forward direction, shows emission characteristics for a selected energy ($\Delta \mathrm{Ene}/A = 1$\,MeV), similar to an RCF layer (summed over front and rear contributions).
  \textbf{Side plots on right:} Energy lineout for pointing sensitivity with a pinhole aperture
  taken for acceptance on axis (black line: $(0\pm 2)$\,$\degree$) and off-axis (grey line: $(7,11,15\pm 2)$\,$\degree$), similar to a typical Thomson parabola.}
\label{fig:a0_scan_emissionDistrib}
\end{figure*}

Figure~\ref{fig:a0_scan_emissionDistrib} shows the angular energy emission distribution in polarization direction.
When comparing ion pointing for varying laser intensities, both hydrogen and deuterium emission distributions are more directed for low $a_0$ than higher $a_0$.
In our setup, front-side protons add a diffuse emission signature on top of the rear-side, directed protons due to hole-boring RPA contributions.
The emission characteristics of hole-boring pre-accelerated ions from the target front surface (e.g. wider or narrower than the TNSA rear and structuring) depends on the given laser contrast and target pre-expansion and requires a detailed parameter scan for a given experiment.
In our specific setup, front-side contributions emit into wider angles than rear-side ions.
Note that the absolute values for energy and emission angle will both be smaller in 3D3V simulations and experiments, an experimental identification for areas of pure front-side contributions might be possible when clear structural changes are visible.

With increasing deuterium content in the target, the visible "hole" (marked red in Figure~\ref{fig:a0_scan_emissionDistrib}) in mid-range protons shifts to higher energies as more deuterium ions can displace the decreasing number of protons more efficiently. 
Experiments will observe a "gap" at low energies and quasi-monoenergetic features when observing strictly along the target normal with narrow angle acceptance.
Under limited cryogenic target orientation stability\cite{Obst2017EfficientJets} such measurements might suffer from shot-to-shot fluctuation of target normal alignment with respect to the fixed diagnostic axis.
The lineouts on the right side of each plot in Figure~\ref{fig:a0_scan_emissionDistrib} exemplify, how a positioning jitter of 7\,$\degree$ ($a_0 = 16$), influences the measured spectra.
Experimental campaigns can mitigate this issue by either collecting enough statistics or deploying wider, angle-resolved spectrometers (resolution $\le 10\,\degree$).

\section{\label{app:target_origin}In-Target Origin of Energetic Ions}

\begin{figure*}[ht]
  \centering
  \includegraphics[width=\textwidth]{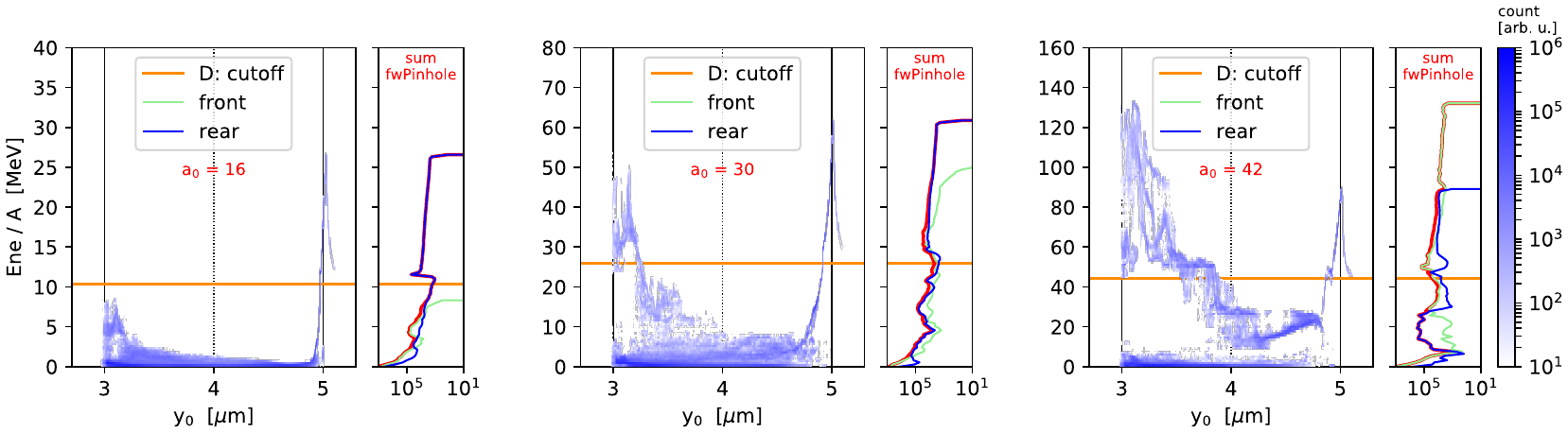}
  \caption{Correlation of final energy depending on initial longitudinal position $y_0$ inside the target for hydrogen ions.
  Plotted for laser intensities $a_0=16,30,42$ and target deuterium ratio $d=90\%$.
  $y_0$ describes the target along the laser-axis (laser from left), target front and rear are defined as initial position $y_0$ before or after $4\,\upmu$m.
  Lines on the right show individual contributions of front (first micron up to $y_0=4$ in the target) and rear ions on the observed energy spectra in "forward pinhole" acceptance.
  }\label{fig:a0_scan_energyOrigin}
\end{figure*}

For TNSA, the highest accelerating fields are expected at the rear surface of the target.
Figure~\ref{fig:a0_scan_energyOrigin} connects the initial target depth of a hydrogen ion with its finally reached energy.
On the right of each figure, the forward-observed proton spectrum (red) with a blue line showing rear and green line showing front side contributions corresponds to the integral over the initial target depth.

Over increased $a_0$, the influence of front-side originating ions onto spectral modulations increases and finally dominates the cutoff energy for the PW-scale $a_0=42$.
Front-side ions are further accelerated in the rear electro-static fields, which unsurprisingly leads to a similar spectral shape.

For an investigated $a_0$ of 30 and above, a deuterium ratio of $90\,\%$ is required to cause a significant modulation in the proton spectra.
Accelerating rear fields scale with laser strengths from increased sheath electron energy.
However, the superimposed multi-species fields that cause the spectral modulation do not increase with $a_0$ and depend on target composition. 

Not only rear-surface protons experience a modulation, but also front-side hole-boring RPA protons are influenced by deuterium ions\cite{Kar2012IonPressure}.
Further details are shown in 
Figure~\ref{fig:a0_42_dratio_0_00_0_80_origin_ypy}.

For $a_0 \gtrsim 30$, at energies in the spectrum above the cutoff of rear-side ions, one might expect a change in slope in the overall proton spectra as front side ions start to dominate the highest energies (red line in Figure~\ref{fig:a0_scan_energyOrigin}).
In some cases, a small energy modulation hints the change between the front and rear contributions which might not be significantly observable, since front-side ions are also pre-dominantly further accelerated in TNSA fields.
As this is the observable obtained in an experiment, it is not possible to quickly infer which part of the spectrum is front- or rear-side originated without further radial emission information.

\section{\label{app:a42_repulsion_front}Multi-Species Effect on Front-Side Ions for $a_0 = 42$}

\begin{figure*}[ht]
  \centering
  \includegraphics[width=0.65\textwidth]{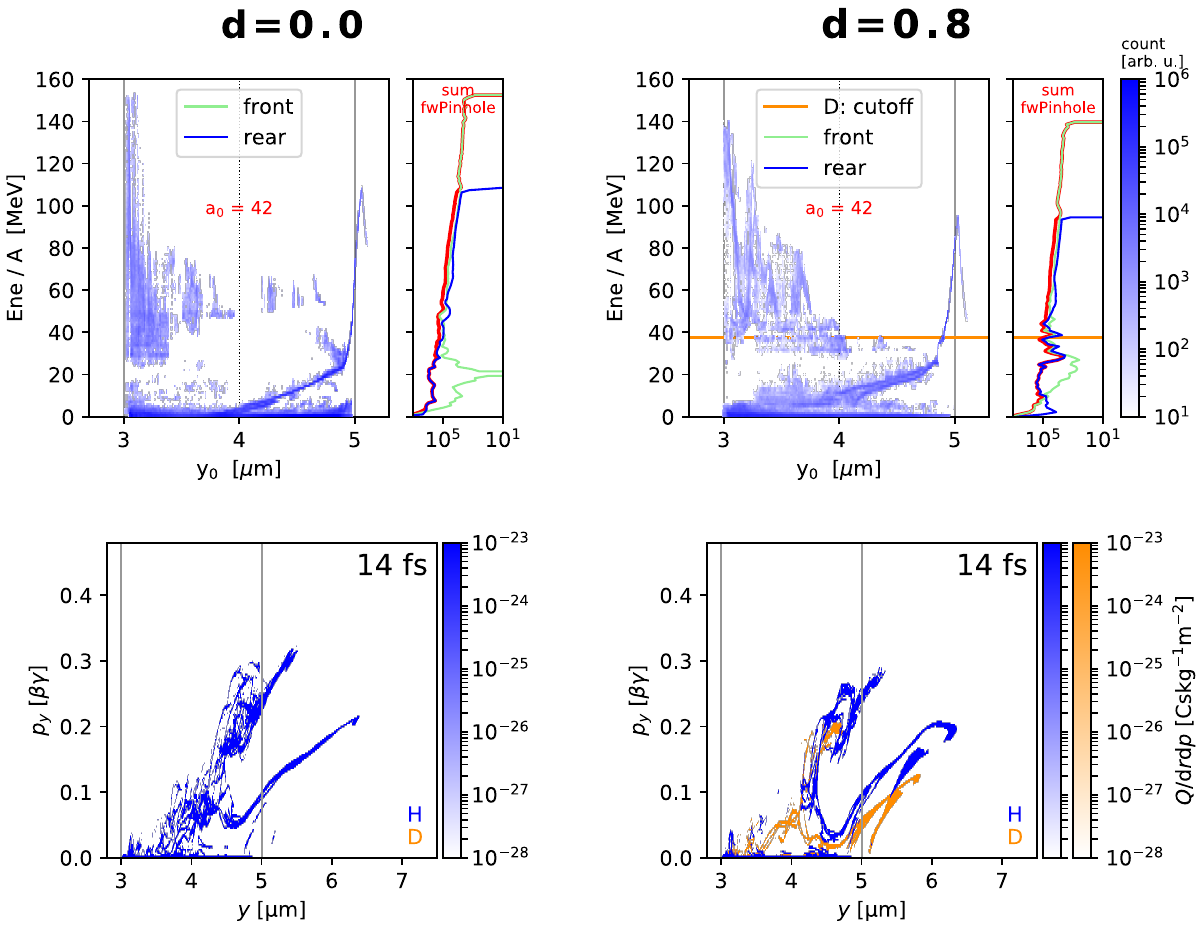}
  \caption{Multi-species effect for front-side ions for $a_0=42$, comparing a pure hydrogen target ($d=0.0$, left column) and a mixed target ($d=0.80$, right column). Selected plots show particles in acceptance for a "forward pinhole". (Upper plots) Origin of final energies for hydrogen ions. (Lower plots) Longitudinal phase space for early times, $14\,$fs after laser peak intensity on target.
  }\label{fig:a0_42_dratio_0_00_0_80_origin_ypy}
\end{figure*}

Figure~\ref{fig:a0_42_dratio_0_00_0_80_origin_ypy} (top) shows a correlation plot between original in-target position and final energy of hydrogen ions for case $a_0=42$.
For the high-deuterium case $d=0.8$, front-side originating ions are equally affected by the multi-species effect from deuterium (see orange line for deuterium cutoff energy).
The reason for that lies in the sufficiently long co-propagation through the target after initial front-side hole-boring acceleration.
The longitudinal phase-space plots (bottom) show once again the proximity of deuterium and hydrogen ions as they are still within the target during burn-through, 14\,fs after peak laser intensity on target.

\section{\label{app:experimental-nodes}Notes for Potential Experimental Realizations}

For significant energy modulation signatures in the high-energy tail of distributions, a deuterium ratio $\ge$~60\,\% is advisable.
The observed multi-species effects are rather robust against small changes in deuterium ratio.
Therefore, the mixing ratio only needs to be generated with a precision of few percent.
Additionally, we observed a increase of proton cutoff energies in simulations for mixing ratios between 20 - 50\,\%. 

For spectrometer setups, an energy resolution for the mid-energy range needs to be chosen that can resolve the few-MeV wide energy modulation.
When deciding for an ion spectrometer with high energy resolution but small angular acceptance such as a Thomson parabola, a statistically significant amount of shots needs to be accumulated in case the target orientation jitter obfuscates the otherwise clear signature in the energy spectrum of target normal directed protons.
The dynamic range in terms of proton counts per energy and solid angle should at least be three orders of magnitude for a good resolution of the start and end of the modulation.

\end{document}